\documentclass[aps,prl,twocolumn,color,psfig,epsf,superscriptaddress]{revtex4-2}
\usepackage[english]{babel}
\usepackage[T1]{fontenc}
\usepackage{tikz}
\usepackage{amssymb,amsmath,amsfonts}
\usepackage{textcomp}
\usepackage{braket}
\usepackage{nicematrix} 
\usepackage{graphicx,color}
\usepackage[utf8x]{inputenc}
\usepackage{bm}
\usepackage{mathbbol}
\usepackage{float}
\usepackage{comment}
\usepackage{siunitx}

\usepackage[linkcolor = blue, citecolor = red, urlcolor = blue, colorlinks = true]{hyperref}
\usepackage{tabularx}

\begin{document}

\newcommand{\bQ}{\mathbf{Q}}
\newcommand{\bn}{\mathbf{n}}
\newcommand{\bv}{\mathbf{v}}
\newcommand{\br}{\mathbf{r}}
\newcommand{\bH}{\mathbf{H}}

\newcommand{\tns}[1]{{\color{pink}[TNS: {#1}]}}

%Spontaneous phase separation and self-assembled smectics in lyotropic liquid crystals
\title{Spontaneous phase separation and pattern formation in a lyotropic    nematic mixture}

% Use letters for affiliations, numbers to show equal authorship (if applicable) and to indicate the corresponding author
\author{A. Bensabat}
\author{O. Skelton}
\author{J. Arlt}
\author{M. Bjelogrlic}
\author{D. Marenduzzo}
\author{G. Negro$^{\dag}$}
\author{T.~N. Shendruk}
\author{T.~A. Wood}

\affiliation{SUPA, School of Physics and Astronomy, Edinburgh University, EH9 3FD, UK\\$^{\dag}$ giuseppe.negro@ed.ac.uk}

\begin{abstract}
Lyotropic liquid crystals can display rich phase behaviour and self-organisation, yet the physical principles underlying their self-assembly into large scale patterns remains understudied. Here, we combine theory, simulations and experiments on Sunset Yellow-water chromonic mixtures to show that such materials spontaneously phase separate, even without assuming any underlying microscopic attraction between the molecular species. In our minimal model, demixing depends solely on the Onsager-like coupling between local nematogen density and orientational order. If such a coupling is sufficiently strong, nematic defects trigger the nucleation of isotropic droplets, which then coalesce due to elastic or interfacial tensions. %to ultimetaly yield macroscopic phase separation or microphase separation, depending on parameters. 
We further show that strong anchoring of the director field at the interface arrests this coarsening process, resulting in a stable microphase separated lamellar pattern. This self-assembled smectic phase has striking and unusual features, including spontaneous undulations, heterogeneous layer spacing, long-lived glassy defect patterns and lamellar onions. Our results identify orientational-density coupling and elastocapillarity as fundamental mechanisms to guide self-assembly in lyotropic and chromonic liquid crystals. 
\end{abstract}

%\dates{This manuscript was compiled on \today}
%\doi{\url{www.pnas.org/cgi/doi/10.1073/pnas.XXXXXXXXXX}}

\maketitle
%\thispagestyle{firststyle}
%\ifthenelse{\boolean{shortarticle}}{\ifthenelse{\boolean{singlecolumn}}{\abscontentformatted}{\abscontent}}{}

%\firstpage[9]{4}

Lyotropic liquid crystals (or simply lyotropics) are composite materials with surprisingly rich phase behaviour and potential for self-assembly~\cite{dierking2020}. Besides being of fundamental scientific interest, they are also appealing technologically, because they can be used as  biosensors~\cite{wang2022}, or as metamaterials with unusual optical, flow or mechanical properties~\cite{saadat2021}. 

Chromonic liquid crystals provide a fascinating example of lyotropics. They consist of molecules that reversibly stack into aggregates stabilised by aromatic interactions~\cite{park2008,lydon2011}. Their biocompatibility together with the ability to form multiple self-assembled phases make them promising candidate vessels for targeted drug delivery biological sensing~\cite{Shiyanovskii01062005}. Sunset Yellow (SSY) is one of the most widely studied chromonics because it readily forms lyotropic nematics and is amenable to experimental investigation~\cite{edwards2008}. Yet, despite this attention, the physical origin of the well-documented broad nematic–isotropic coexistence region in SSY–water mixtures ~\cite{janowitz2005,joshi2009,edwards2008,park2008,nastishin2005} remains poorly understood. In particular, there is little evidence that SSY molecules experience strong mutual attraction, raising the question of what controls the observed phase separation.

From a fundamental physics viewpoint, mixtures of liquid crystalline and isotropic fluids exhibit strikingly non-trivial emergent behaviour~\cite{matsuyama2002,lintuvuori2018}, due to the  interplay between elasticity, anchoring and interfacial stresses. This behaviour can be captured by several key dimensionless numbers, such as the elastocapillary number determining the balance between elastic and interfacial stress~\cite{lintuvuori2018}, and the ratio between anchoring strength and surface tension~\cite{meister1996}. The balance between anchoring strength and elasticity is also well-known to determine the defect structure around inclusions~\cite{stark2001,Musevich2006,Smalyukh2018,LAVRENTOVICH201697,Keefe2025,Negro2025}, or the self-assembly of colloids and defects at air-water interfaces~\cite{lintuvuori2013}. The diversity of patterns and richness in behaviour observed across lyotropics arises because these dimensionless ratios can take on a wide range of values, corresponding to different balances between the associated forces, and hence to different physical regimes. %according to the system parameters.

In this work, we combine computer simulations and theory to study phase separation and pattern formation in a lyotropic system; we also compare the predicted phase behaviour with that we observe experimentally in mixtures of SSY and water at different compositions and temperatures. The theoretical free energy functional which we use  favours the mixed phase in the absence of liquid crystalline order -- in other words, we do not assume any attractive interactions between the molecular species which would lead to demixing. First, such a minimal model for a lyotropic mixture is useful because the results are likely to apply generally, to a large class of materials. Second, this premise is a reasonable simple starting point for SSY, where the constituent molecules have charged sulfonate groups at their periphery, leading to electrostatic repulsion between self-assembled SSY stacks, so that any effective like-charge attraction between stacks would likely need to result from more complicated fluctuation-induced many-body interactions (see the discussion in~\cite{park2008}).
An important result we obtain with our minimal model is that %, notwithstanding our assumption of no underlying molecular attraction, we nevertheless find that 
the Onsager-like coupling between local nematogen density and orientational order favours the formation of isotropic droplets at nematic defects, which nucleates spontaneous phase separation. This provides a robust mechanism to rationalise the generic presence of a sizeable phase separation region in the phase diagram of SSY, without the need to assume any significant affinity between SSY stacks.

Our theory also predicts that coarsening is arrested and the system forms a stable lamellar system when the anchoring of the nematic director at interfaces -- either tangential or normal -- is sufficiently strong. This supramolecular self-assembled smectic (or ``super-smectic'') phase is also observed experimentally in our chromonic SSY mixture. Unlike conventional smectics, this self-assembled structure is characterised by pronounced undulations, which lead to the formation of long-lived, and possibly glassy, defect patterns, including edge dislocations and lamellar onions. Whilst the lamellar size is self-selected and well-defined, the spacing between layers is heterogeneous and uneven, suggesting that the layer-compression modulus of our self-assembled smectics is unusually small. This property may be the reason why onions form in our system, in contrast with usual smectics where these patterns are rare, especially in the absence of a shear flow~\cite{ramos2004}. Additionally, the fact that our super-smectic contains essentially frozen defect patterns which depend on initial condition and sample history suggest that lyotropic mixture such as SSY may provide an understudied example of multistable self-assembled and tunable glass with biocompatible properties.

\begin{figure*}%[tbhp]
    \centering
    \includegraphics[width=.8\linewidth]{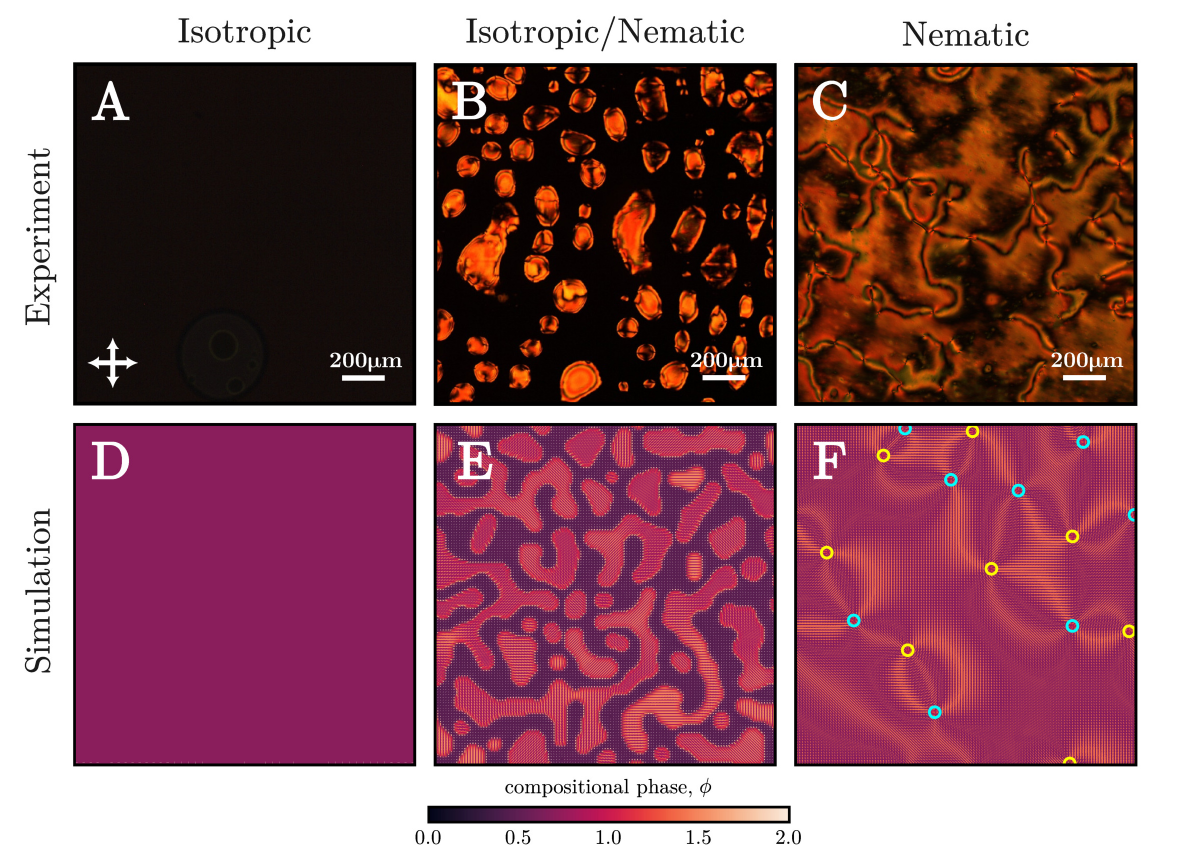}
    \caption{\textbf{Simulations replicate the isotropic–nematic (I–N) transition and phase separation.}
    (A–C) Experimental images showing the isotropic (A), coexistence (B), and nematic (C) phases as the temperature decreases ($\gamma_0$ increases) and the nematic fluid concentration $\phi_0$ increases. (D–F) Corresponding simulation snapshots illustrating the same sequence of phase behaviour for a 128x128 system. The colour map represents the local compositional phase, $\phi$, while the overlaid lines indicate the nematic director field, with line length proportional to the local degree of orientational order. The cyan and yellow rings represent topological defects of charge +1/2 and -1/2, respectively. Scale bars: \SI{200}{\micro\meter}. Vertical and horizontal arrows in panel A show the polariser and analyser orientations, respectively.}
    \label{fig:snapshots}
\end{figure*}

\begin{figure*}%[tbhp]
    \centering
    \includegraphics[width=\linewidth]{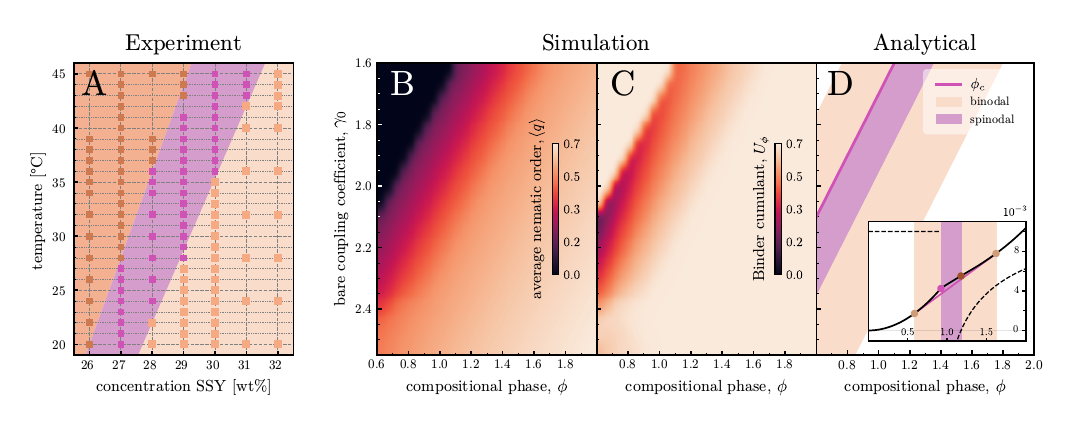}
    \caption{\textbf{Experimental, simulated, and analytical phase diagrams reveal the isotropic–nematic transition and coexistence regions.}
     (A) Experimental phase diagram of SSY solutions showing isotropic (brown), coexistence (violet), and nematic (beige) regions as functions of concentration and temperature. (B and C) Simulated phase behaviour represented by the average nematic order $\langle q \rangle$ and Binder cumulant, $U_{\phi}$, as functions of bare coupling coefficient $\gamma_0$ and composition $\phi$. (D) Analytical phase diagram derived from the common tangent construction on the free-energy density $f_{\text{hom}}$, showing the binodal and spinodal limits. The pink line denotes $\phi_c$, the nematic fluid concentration marking the transition from a homogeneous to a phase-separated configuration. Inset: Schematic of the common tangent construction, plotting the homogeneous free energy density $f_{\text{hom}}$ (solid line) and its second derivative $\partial^2_{\phi} f_{\text{hom}}$ (dashed lines) as a function of $\phi$. The tangent points (beige circles, $\phi_\pm$) define binodal limits, while regions of negative curvature define spinodal boundaries (bounded by the pink circle, denoting the discontinuity in $\partial^2_{\phi} f_{\text{hom}}$, and the brown circle), indicating that phase-separated states are energetically favoured over homogeneous ones.}
    \label{fig:phase_diagrams}
\end{figure*}

\section*{Spontaneous Phase Separation in Lyotropics}

\subsection*{Experimental and simulation phase diagrams for lyotropics show broad coexistence regions}

We start by reporting results on the experimental phase behaviour of SSY-water mixture (see Materials and Methods for details of materials preparation, and free energy functional used in the simulations). Our experiments show that, by changing composition and temperature, three regimes can be readily observed: an isotropic phase (Fig.~\ref{fig:snapshots}A), a nematic phase (Fig.~\ref{fig:snapshots}C), and an intermediate phase separated regime, where isotropic and nematic domains coexist (Fig.~\ref{fig:snapshots}B). This coexistence region is striking because SSY monomers should in principle exhibit electrostatic repulsion between them due to the sulfonate groups~\cite{nastishin2005} suggesting that a different mechanism other than an underlying thermodynamic attraction between SSY stacks may drive demixing.

%\textcolor{red}{AB: Furthermore, from my understanding, the cited work seems to suggest that the increase in concentration leads to longer (and more stable) molecule stacks which in turn favours sulfonate-water attraction which leads to a formation of droplets. I'm not sure we can argue this.}

\begin{figure*}%[tbhp]
    \centering
    \includegraphics[width=0.85\linewidth]{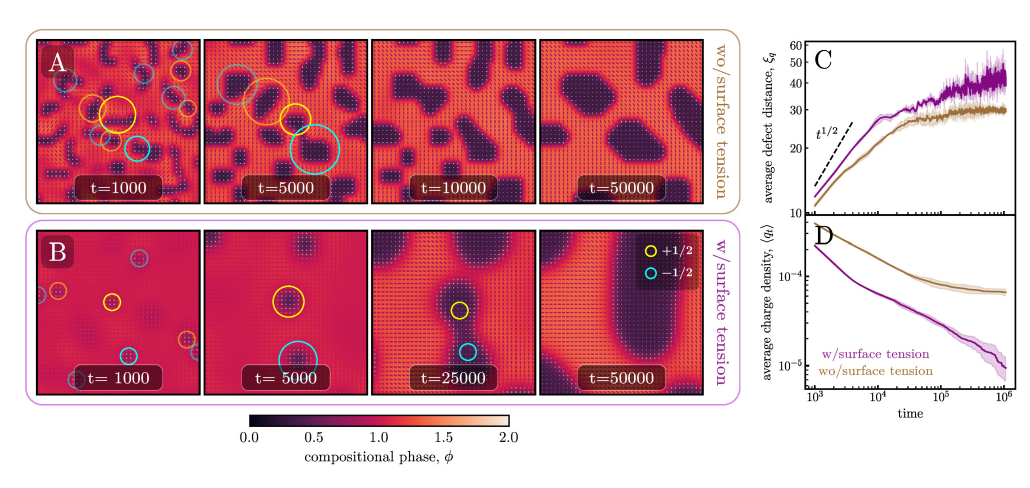}
    \caption{\textbf{Defect-driven phase separation and droplet coarsening dynamics with and without surface tension.}
    Starting from a random, noisy initial configuration, the nematic field evolves toward macroscopic phase separation through the annihilation of +1/2 (yellow) and -1/2 (blue) topological defects, leading to the formation of isotropic voids that grow and coarsen over time. Panel A shows a simulation without surface tension ($\kappa=0.00$) and panel B includes surface tension ($\kappa=0.01$). The column on the right shows the the average defect distance over time, $\xi_q$ (C) and the average charge density, $q_t$ (D). Both quantities are averaged over three runs with different randomised initial configurations in which the mean compositional phase is fixed at $\phi_0=1$ and the bare coupling coefficient at $\gamma_0=2.1$. Both B and C show magnified views ($40 \times 40$) of the global simulation domain ($128 \times 128$). The full animations are available in Suppl. Movies 1 and 2. A qualitatively similar coarsening is observed experimentally when a mixture of SSY at 28 wt$\%$ undergoes a temperature ramp (Suppl. Fig. 2 and Suppl. Movie 3).}
    \label{fig:domain_growth}
\end{figure*}

Remarkably, even our minimal model for lyotropic liquid crystals (see Materials and Methods), which contains no anchoring term, no explicit interfacial tension, and no attractive interactions, reproduces all three regimes observed experimentally (Fig.~\ref{fig:snapshots}D-F). In the phase separated region, the system self organises into a number of nematic rafts which are not circular, suggesting that elasticity leads to the emergence of an effective weak, or anisotropic, surface tension. In the simulations, the rafts exhibit sharper features due to the omission of explicit interfacial tension. Reintroducing it results in smoother, rounded morphologies that more closely match experimental observations (Fig.~\ref{fig:domain_growth}A–B). These nematic rafts tend to be internally ordered and defect-free in simulations (Fig.~\ref{fig:snapshots}E); experimentally, some defects remain as apparent from the Schlieren patterns in Fig.~\ref{fig:snapshots}B, possibly reflecting the presence of some non-negligible anchoring in the experiments. %\tns{can we make the same claim about experiments?}~\textcolor{red}{AB: In the experiments they seem to have some defects, but that system is not entirely 2D. It seems to me that the Schlieren patters still show some defects.}
Defects are instead transiently observed in simulations in the nematic regime (Fig.~\ref{fig:snapshots}F), which is qualitatively consistent with experimental observations (Fig.~\ref{fig:snapshots}C). 

%describe experimental and simulation phase diagram
More quantitatively, we systematically map the phase diagram of our chromonic SSY mixture in our experiments for different values of composition ($26-32$ wt \%) and temperature ($20-45^{\circ}$C). The phase diagram features linear phase boundaries and a broad coexistence region (Fig.~\ref{fig:phase_diagrams}A). The phase diagram is in good agreement with those found in the literature for SSY mixtures~\cite{edwards2008,janowitz2005,joshi2009}. Within the coexistence region, the ratio of nematic to isotropic domains changes with temperature, an example of which is shown in Suppl. Fig. 3.

Our minimal model for lyotropics leads to a phase diagram which is similar in shape (Fig.~\ref{fig:phase_diagrams}B,C), where phases are identified by measuring the average nematic order and the Binder cumulants computed from the probability distribution functions for local composition $\phi$ (see SI). %~\cite{SI}\tns{cite SI?}. 
Unlike conventional LC mixtures, which show curved binodals typical of attraction-driven demixing~\cite{matsuyama2002,araki2004,sulaiman2006,dagama2025}, both the experimental SSY–water mixture and the minimal model exhibit nearly linear phase boundaries, suggesting that these mixtures are qualitatively different from those considered theoretically in the past, and that the functional form employed for the free energy potential in our model is the appropriate one for our experimental system. %coexistence originates from a different mechanism than classical mixing thermodynamics. %the orientational–density coupling rather than from classical mixing thermodynamics. %We therefore hypothesise that these linear boundaries are the signature of a different type of phase separation, which do not require underlying molecular attraction, and which we refer to as spontaneous phase separation. 

\subsection*{Coupling-Induced Phase Separation}

To explain the phase diagram observed in both experiments and simulations, we analyse the equilibrium thermodynamics of the binary mixture. %As established in previous work~\cite{AssanteEtAl2023}, 
Within our model, the bulk free energy is coupled to the composition $\phi$ through an effective inverse reduced temperature $\gamma(\phi) = \gamma_0+\Delta \ \phi$, where $\Delta$ is the coupling parameter. 
The uncoupled system ($\Delta=0$) mixes freely implying that the phase separation observed here is driven entirely by the coupling between the composition, $\phi$, and the nematic order, whose magnitude is $q$.

We determine the phase boundaries using the common tangent construction~\cite{AssanteEtAl2023}, applied to the free energy of a homogeneous configuration of unvarying composition and uniformly aligned nematic phase. In this state, the elastic term vanishes (see Eq.~\eqref{eq:f_LC}), and the homogeneous free energy density depends only on the nematic order $q$ and the composition $\phi$ as
\begin{equation}
    \begin{split}
           f_\text{hom}(q,\phi) = & \frac{A_0}{3}\left(1-\frac{\gamma}{3}\right)q^2 - \frac{2 A_0\gamma}{27}q^3 \\
           & +\frac{A_0\gamma}{9}q^4+\frac{a}{2}\phi^2.
    \end{split}
    \label{eq:f_hom}
\end{equation}
In this homogeneous configuration, the local nematic density is constant in space, $\phi(\br)=\langle\phi\rangle$ for all locations $\br$. The total free energy is thus minimised for the nematic order
\begin{equation} 
    q_\text{min} (\phi) = 
    \begin{cases}
        0, & \text{if } \gamma\leq \gamma_c \\ 
        \frac{1}{4}\left(1+\sqrt{9-24/\gamma(\phi)}\right), & \text{if } \gamma\geq\gamma_c.
    \end{cases} 
    \label{eq:q_min}
\end{equation}
The homogeneous configuration is stable against phase decomposition only if the corresponding free energy, $f_\text{hom}(q_{\text{min}},\phi)$, is convex. However, when coupling between compositional and orientational order is considered, the free energy can become concave, defining a spinodal region where the second derivative of the free energy is negative and the mixture is unstable (Fig.~\ref{fig:phase_diagrams}D; inset). The stable coexisting phases (binodals) are identified via the common tangent construction using an iterative method to find the unique tangent line at two distinct concentrations, $\phi_-$ and $\phi_+$. This line represents the average free energy density of the phase-separated mixture, which is lower than that of the homogeneous state for any given average composition $\langle \phi \rangle$ lying between the binodals.
The spinodal region determined semi-analytically is in agreement with the phase separation regions mapped in our experiments and simulations (Fig.~\ref{fig:phase_diagrams}); it also compares well with that found in other experimental works~\cite{edwards2008,janowitz2005,joshi2009}.

\subsection*{Elastic and interfacial stresses give different demixing dynamics} %{Isotropic droplet formation at nematic defects nucleate phase separation}

Whilst the previous theory indicates whether or not the system is unstable to demixing, it does not reveal how phase separation proceeds dynamically, neither does it predict whether the end result is macroscopic phase separation or microphase separation. To address these two questions, we have used simulations %and experiments 
to follow the demixing dynamics (Fig.~\ref{fig:domain_growth}).

%\textcolor{red}{AB: Missing description of the experimental snapshots and comparison to simulations? Currently the experimental snapshots are from a temperature ramp and not for a demixing as in our case.}

We first consider the case of a lyotropic mixture with negligible interfacial tension ($\sigma=0$) between the two fluids (Fig.~\ref{fig:domain_growth}A; Methods). Starting from a noisy nematic phase, nucleation of defects with topological charge $\pm 1/2$ occurs. Instead of annihilating, as would occur in a single-phase nematic fluid, these defects first nucleate a local change in composition resulting in the formation of small isotropic voids where the defects were (Fig.~\ref{fig:domain_growth}A). As the nematic director near the droplet interface retains memory of the defect patterns, elastic stresses drive the coalescence of neighbouring droplets which have opposite value of local topological charge. 
As in a single-phase nematic~\cite{chuang1991}, the number of defects, or more precisely the integral of the modulus of the topological charge density (see Methods) decreases over time, because such defects introduce an elastic stress in the material which is proportional to the elastic constant $K$. The decrease in number of defects causes the average distance between defects to increase with the square root of time, $\xi_q \sim t^{1/2}$ (Fig.~\ref{fig:domain_growth}C), following the scaling law governing defect coarsening in single-phase nematics~\cite{chuang1991}. Concomitantly, we observe that the average charge density $\left\langle q_t\right\rangle$ decreases. 
In the lyotropic mixture, this topological charge annihilation process drives droplet coalescence, and because of the decrease in the overall topological charge with time, this coalescence-based coarsening slows down dramatically and appears to arrest at late times, with the average charge density plateauing to small non-zero values at long times (Fig.~\ref{fig:domain_growth}D). 
This kinetic arrest explains the microphase separation observed in our simulations for negligible surface tension.

For sufficiently strong interfacial tension $\sigma$, the growth of isotropic domains proceeds in a qualitatively different way (Fig.~\ref{fig:domain_growth}B). The driving force for coalescence is now provided by both elasticity and the interfacial tension; as a result, coarsening proceeds indefinitely (Fig.~\ref{fig:domain_growth}C-D), leading to macroscopic phase separation in steady state. 

We suggest that the dynamical crossover between elastically-driven-then-arrested coarsening and interfacially-driven macrophase separation is controlled by the elastocapillary number $\text{Ec}$ controlling the balance between interfacial and elastic stresses~\cite{lintuvuori2018}, which in our minimal model can be written simply as $\text{Ec}=\kappa/K$, where $\kappa$ is related to the interfacial energy and $K$ is the nematic elastic constant (see Methods). When $\text{Ec}$ is small, coarsening is driven by elasticity and interaction between local residual topological charges, whereas coarsening is driven by surface tension and resembles standard binary fluid phase separation when $\text{Ec}$ is large~\cite{bray1994}. 
The dimensionless elastocapillary number is $\text{Ec}=0$ in Fig.~\ref{fig:domain_growth}A and $\text{Ec}=1$ in Fig.~\ref{fig:domain_growth}B. %\textcolor{red}{AB: It could be me, but it doesn't seem direct to get an analytical expression for the surface tension because we don't have an explicit double well}. 

%ANA: I agree that we can use this dimensionless number. DAVIDE: Perfect!
% :-)

\section*{Pattern Formation in Lyotropic mixtures}

\begin{figure*}[tbhp]
    \centering
    \includegraphics[width=1.5\columnwidth]{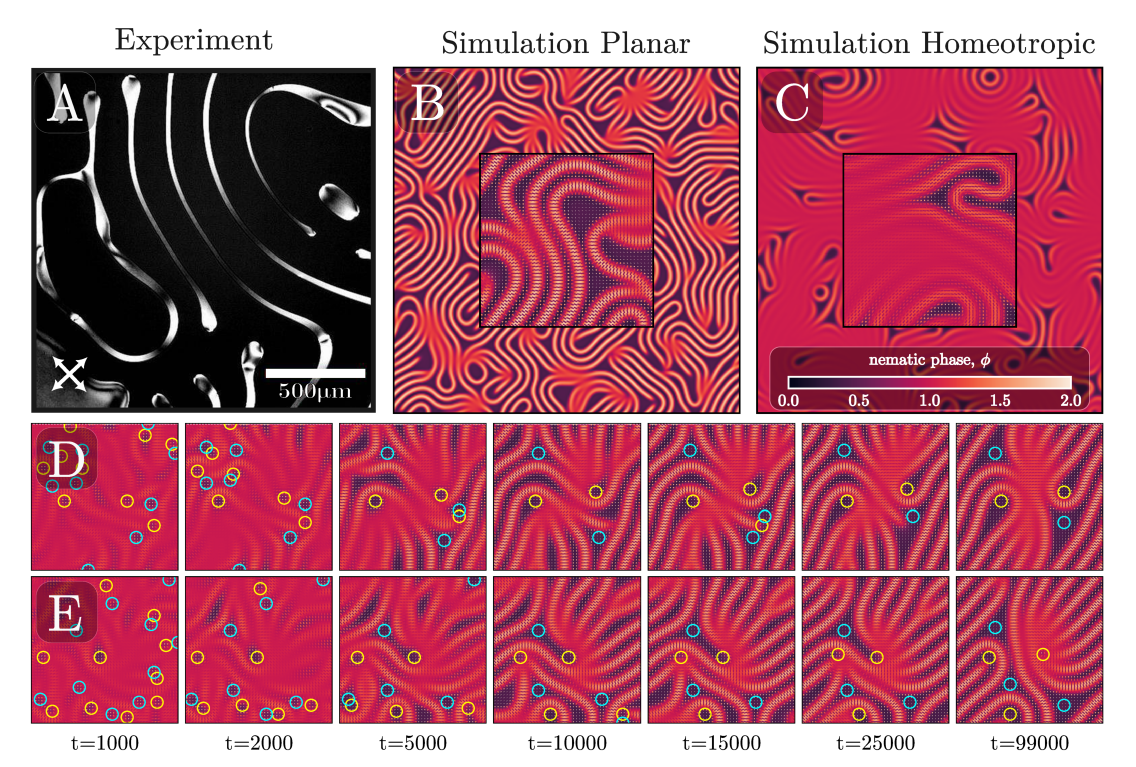}
    \caption{\textbf{Anchoring-induced microphase separation and lamellar ordering in the isotropic–nematic mixture.}
    (A) Experimental snapshot showing phase separation into alternating nematic and isotropic layers (see Suppl. Movie 4 for the complete dynamics). (B-C) Corresponding simulations exhibiting similar lamellar structures, albeit more densely packed. Simulation parameters are $\gamma_0=2.0$, $\phi_0=1.0$, and anchoring strength $W=-0.03$ (B), corresponding to normal anchoring, and $W=0.06$ (C), corresponding to planar anchoring. (D-E) Time evolution of two simulations with different initial conditions (same parameters as in B) showing defect dynamics: $+1/2$ (yellow) and $-1/2$ (blue) defects appear, move, and annihilate to relax the lamellar structure (see also Suppl. Movies 5 -6). In some cases, defects reappear transiently (D), while in others, a persistent $+1/2$ defect acts as a source of nematic layers (E). In both cases, complete annihilation is inhibited, leading to arrested, anchoring-stabilised microphase separation. Lower left to top right and top left to lower right arrows in panel A show the polariser and analyser orientations, respectively.}
    \label{fig:super_smectics}
\end{figure*}

%XXX caption of Fig. 4 needs changing as no self-correlations

\subsection*{Anchoring induces the self-assembly of a supramolecular smectic phase}

The results discussed up to now can be understood within a model which does not include any interfacial anchoring of the director field. Introducing anchoring -- either normal or tangential -- leads to additional physics and pattern formation (Fig.~\ref{fig:super_smectics} and Suppl. Movies 5-6). 

In particular, as the anchoring strength $W$ increases past a critical threshold, a completely different phase self-assembles starting from a uniform disordered phase with noise. This is a self-assembled lamellar phase that is observed experimentally (Fig.~\ref{fig:super_smectics}A and Suppl. Movie 4), close to the transition from coexistence to the isotropic phase. This lamellar phase is also observed in simulations (Fig.~\ref{fig:super_smectics}B). 

This lamellar phase is a supramolecular smectic: the domains consist of elongated layers of  microphase separated nematic domains, as opposed to molecular layers. 
We therefore refer to this phase as a self-assembled ``super-smectic''.  The difference with respect to lamellar phases in lyotropics and copolymers~\cite{morozov2001} is that this super-smectic phase is stabilised by anchoring along the interfaces of the layer-like microphase separated domains.

%lamellar phases in lyotropics, is this super-smectic different?

%\subsection*{Anchoring leads to a Negative Effective Surface Tension}

\begin{figure}%[tbhp]
    \centering
    \includegraphics[width=\linewidth]{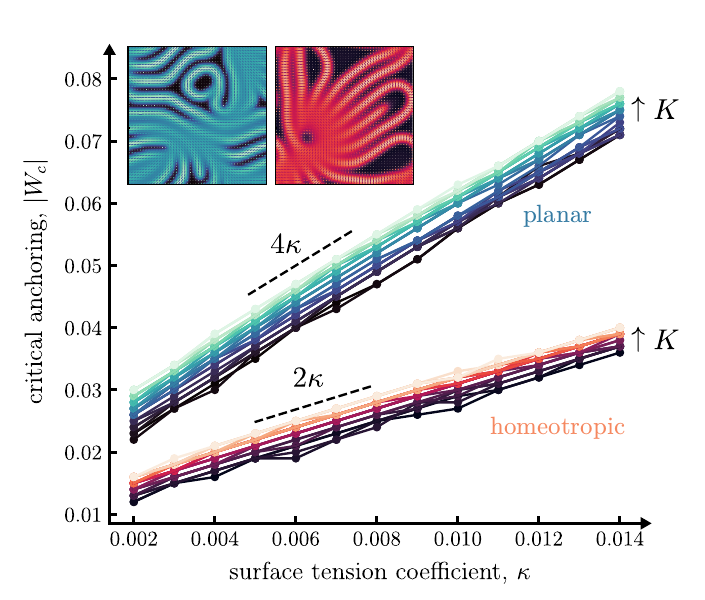}
    \caption{\textbf{Critical anchoring for planar and homeotropic cases.} The critical anchoring follows the theoretical predictions (dashed lines) for both planar (blue) and homeotropic (orange) anchoring as a function of the surface tension coefficient, $\kappa$. Line brightness increases as a function of the elastic nematic constant, $K$, varied from 0.10 to 0.30. The insets on the top left show typical system configurations at critical anchoring strength $|W_c|$.}
    \label{fig:critical_anchoring}
\end{figure}

To understand more quantitatively the mechanism driving the self-assembly of a super-smectic, we examine the thermodynamic stability of a single lamellar nematic domain with a well-defined interface. We consider a region where the director field $\mathbf{n}(\mathbf{r})$ is spatially uniform and coupled to the domain interface. 
The director $\mathbf{n}$ can either be parallel to the layer normal direction $\nabla \phi$ (homeotropic anchoring) or perpendicular to $\nabla \phi$ (planar anchoring). Assuming that the local nematic order is linearly coupled to the local concentration via $q(\mathbf{r}) \approx \alpha\phi(\mathbf{r})$, the free energy density can be recast to isolate the gradient contributions
\begin{equation}
\begin{split}
f_\text{lam} \approx f_\text{bulk}(\phi, q) + \kappa_\text{eff}(\phi, q) (\nabla \phi)^2,
\end{split}
\label{eq:f_lam}
\end{equation}
where an effective surface tension coefficient
\begin{equation}
    \kappa_\text{eff} = \frac{\kappa}{2} - \frac{W q}{3} + \frac{K \alpha^2}{3} + Wq \frac{\bn \cdot \nabla \phi}{|\nabla \phi|}
    \label{eq:sigma_eff}
\end{equation}
appears. The last term in $\kappa_\text{eff}$ captures an asymmetry between homeotropic and planar critical anchoring, which is observed in the simulations (Fig.~\ref{fig:critical_anchoring}). 

Macrophase separation is driven by a positive surface tension ($\kappa_{\text{eff}} > 0$), which drives the system to minimise interfacial area. However, the second term in $\kappa_{\text{eff}}$ (Eq.~\eqref{eq:sigma_eff}) is negative and so a transition occurs when this anchoring contribution renders $\kappa_{\text{eff}}$ negative. In this regime, the creation of interfaces is in fact favoured: the mechanism leading to microphase separation is therefore effectively the same as that of Brazovskii theory~\cite{brazovskii1975,gonnella1997,Xu2006}, although in our case it is the anchoring which renders the surface tension effectively negative. 

In the super-smectic phase, maximising interface contour length whilst minimising surface tension is achieved via long stacked lamellae. The presence of the lamellae interrupts pathways to the global free energy minimum, which arrests the coarsening process and stabilises smaller length scales characteristic of microphase separation. The critical threshold for this transition ($\kappa_{\text{eff}} = 0$) is defined by
\begin{equation}
    \left|W_c^\perp\right| = \frac{3}{2q}\left(\frac{\kappa}{2}+\frac{K\alpha^2}{3}\right), \quad
    % W_c^\parallel = \frac{3}{q}\left(\frac{\kappa}{2}+\frac{K\alpha^2}{3}\right)
    W_c^\parallel = 2 \left|W_c^\perp\right|
    \label{eq:Wc_analytical}
\end{equation}
for homeotropic and planar anchoring, respectively. Evaluating these expressions at the center of the interface between the nematic \textit{lamellae} and the surrounding isotropic fluid makes a quantitative prediction for the critical values. We approximate the local composition as the midpoint between the nematic phase ($\phi \approx 1.5$ in the simulations) and the isotropic fluid ($\phi \approx 0.5$ in the simulations), yielding $\phi \approx 1$. As such, assuming linear coupling, the nematic order at the interface is approximately half the magnitude in the ordered phase. A typical value for inner lamellar structures found in the simulations of $q_\text{lam} \approx 2/3$ (implying $\alpha \approx 0.37$), predicts the characteristic interface value $q_c =\alpha \phi_\text{interface}\approx 0.37$. Substituting these values into Eq.~\eqref{eq:Wc_analytical} yields $|W_c^\perp| \approx 2\kappa + 0.2 K$.
%$|W_c^\perp| \approx 2.027\kappa + 0.185 K$.

This prediction is validated by exploring phase space via numerical simulations. The transition to microphase separation is identified by the emergence of Bragg peaks in the structure factor at wavenumbers corresponding to a characteristic lamellar width $\lambda^* \approx 6$~(Suppl. Fig. 5). The critical anchoring is seen to be linear with $\kappa$, increasing with $K$ and twice as large for parallel anchoring (Fig.~\ref{fig:critical_anchoring}). A linear regression of the simulated critical thresholds yields
\begin{subequations}
    \begin{align}
         |W_c^\perp|_{\text{sim}} &=  2.001\kappa + 0.138 K + 0.008, \\
         |W_c^\parallel|_{\text{sim}} &= 3.987\kappa + 0.253 K + 0.014,
    \end{align}
    \label{eq:Wc_fit}
\end{subequations}
with a fit uncertainty of $\pm 0.001$ (enforcing zero intercepts leads to similar slopes but with a larger error, see SI). %\tns{How bad is the fit if you enforce zero intercept. I guess not great}\textcolor{red}{(I have put it in the SI. it is worse, but I think we are okay even without forcing the intercept)}. 
These results are in good agreement with the analytical estimate, particularly regarding the dominant $\kappa$ contribution, where the simulations reproduce the predicted asymmetry and recover the factor of two difference between the regimes. Deviations in the $K$ dependence may be attributed to simplifications in the model: the linear coupling approximation within the steep gradients of the interface, and our focus on a single lamellar domain, when the entire system exhibits a more complex texture where curvature of the lamellae (splay and bend of the director field) are present. Despite these differences, the overall trends support the proposed anchoring-driven mechanism of microphase separation.

\subsection*{Elastic stresses template the formation of glassy defect patterns}

A notable feature of the self-assembled super-smectic is that in it there are several undulations and long-lived defects in the lamellar ordering. A general mechanism to create undulations in layered systems -- such as smectics or cholesterics -- is the Helfrich-Herault instability, which creates deformations in response to geometric frustration, for instance arising from dilations or compression with respect to the thermodynamically favoured spacing~\cite{blanc2023}. 

However, this is unlikely to be the responsible mechanism of undulations in the super-smectic phase, as initialising the system with gaps between uniformly spaced layers leads to highly heterogeneous patterns in steady state, suggesting that there is no thermodynamic driving towards even layering (Suppl. Movie 7 and 8). In other words, the super-smectic phase has an unusually small compression modulus, so that it should not undergo an Helfrich-Herault instability. Although the system does not favour evenly spaced layering, the layer width itself is self-selected and well-defined, as configurations initialised with different widths evolve towards a characteristic thickness (Suppl. Movie 9).

Inspection of the kinetic pathways through which the undulating patterns form (Suppl. Movies 7, 8 and 9) suggests that pattern selection is instead strongly dependent on initial random seeding due to the initial defect dynamics. 
The defects, which are formed early on, due to the noise in the initial configuration, result in elongated voids as isotropic droplets with opposite effective topological charge annihilate. Charged droplets which do not annihilate nucleate defects, such as smectic disclinations %\tns{Are we sure these are disclinations, not dislocations?}\textcolor{red}{I think so, since the director rotates around the void/defect.} 
(Fig.~\ref{fig:critical_anchoring}; red inset) or lamellar onions (Fig.~\ref{fig:critical_anchoring}; blue inset). Following formation, such defects are kinetically arrested against coalescence, as coalescence would require large scale layer rearrangement, or tunnelling of voids through layers. The spontaneous formation of onion patterns is remarkable, as in usual smectic phase these patterns normally require spontaneous curvature~\cite{ramos2004} or shear~\cite{vanderlinden1996}. 

%could check free energy

\section*{Discussion and Conclusion}

In summary, we have combined theory, simulations and experiments to study the phase separation and pattern formation of a lyotropic liquid crystal composed of a mixture of a nematic and an isotropic fluid, under conditions for which there is no known underlying molecular attraction between the species promoting demixing. On one hand, this minimal model is expected to be broadly applicable to lyotropic nematic mixture. On the other hand, it provides a meaningful starting point to analyse our experimental system, the well-studied chromonic SSY-water mixture, which features a broad phase coexistence region in the phase diagram in the absence of a well-defined mechanism to promote attraction between self-assembled stacks of SSY molecules~\cite{park2008}. Whilst chromonics are notorious for their low twist elastic constant, twist does not appear to play a role in our quasi-2D geometry, hence our minimal single elastic constant approximation is reasonable in this context. 
% Therefore the results for our minimal model are relevant for this experimental system, which is a model lyotropic. 

Our simulations and theory show that phase separation can emerge in lyotropics purely due to the coupling between local nematogen density and orientational order. This provides a robust physical mechanism to explain the large region of phase coexistence which is experimentally observed.
Essentially, demixing arises because forming an isotropic droplet in a nematic domain is, in a large region of parameter space, more favourable thermodynamically than defect formation. 

The kinetic pathway accompanying demixing and the growth of nematic and isotropic domains depends on the balance between elasticity and interfacial stresses, which is controlled by an elastocapillary number $\text{Ec}$. 
% which determines the relative importance of elastic versus interfacial stresses.
When elasticity dominates (for small $\text{Ec}$), demixing is mainly driven by the coarsening of topological defects in the nematic phase, which is observed in both experiments and simulations. 
% as the system is cooled from high to low temperature. 
Defects nucleate voids, which become isotropic droplets inside nematic domains. As mentioned above, these voids are less costly thermodynamically than the elastic deformations near defect cores, however these droplets retain a net topological charge, which drives (through elasticity) coarsening and merging of  droplets with opposite sign of the topological charge.
Coalescence arrests once the remaining droplets have negligible topological charge. 
If, instead, the surface tension between the two components of the lyotropic mixture is sufficiently high (for large $\text{Ec}$), coarsening proceeds indefinitely, resulting in macroscopic phase separation, in a similar manner to model B or model H dynamics, which describe phase separation in binary mixtures of isotropic fluids~\cite{chaikin1995,cates2018}.

%additionally there is an unexplained puzzle in the physics of SSY, that the size of stacks at least as measured with some methods is apparently not large enough to give nematic order with Onsager theory, see \cite{park2008}; can we say something on this from our model?  

A notable prediction of our minimal model, which is verified experimentally in SSY-water lyotropic mixtures, is the self-assembly of a supramolecular lamellar phase composed of nematically ordered microdomains, or a super-smectic. Anchoring provides a general and robust mechanism to self-assemble a super-smectic phase, because it acts as a surfactant thermodynamically, leading to an effectively negative surface tension which drives microphase separation. 
The super-smectic phase is observed experimentally near the transition to the isotropic mixed phase. 

Both experiments and simulations show that the lyotropic super-smectic phase is unusual, in that it contains multiple undulations and defects, such as lamellar onions. Undulations and onions normally appear due to shear~\cite{vanderlinden1996,arrault1999}, geometric frustration~\cite{blanc2023}, or spontaneous layer curvature~\cite{ramos2004}. In our case, these defects are seeded by the topologically charged voids arising early on during the dynamics, which, once trapped in the gaps between the nematic layers, lose mobility leading to kinetically frozen or glassy defect patterns. This mechanism of formation substantially enhances the pattern formation potential of these lyotropic materials, as different initial conditions can in principle seed and template the formation of different glassy defect profiles. This is interesting for future applications, especially as our experimental formulation of this material in terms of chromonic SSY is both widely available and biocompatible~\cite{chavda2023}, raising the possibility of building multistable soft glasses via sustainable self-assembly. 

\section{Methods}
This section presents the hydrodynamic model used in this work. The dynamics of the lyotropic nematic mixture are described by three coarse-grained fields: (i) the fluid velocity $\bv(\br,t)$, (ii) the tensor order parameter $\bQ(\br,t)$~\cite{deGennesProst1993} describing the orientational order of the nematic liquid crystal, and (iii) the scalar order parameter $\phi(\br,t)$ representing the mixture composition.
Furthermore, we detail the preparation of the SSY lyotropic phase, as well as the methods used to construct the phase diagram and perform the experiments.

\subsection*{Nematic Fluid}

The nematic liquid crystal is described by a tensor order parameter $\bQ$ that encodes the alignment of rod-like molecules in the liquid.  In the uniaxial approximation, $\mathbf{Q}$ is a traceless symmetric tensor of the form $Q_{\alpha\beta}=q(n_\alpha n_\beta-\delta_{\alpha\beta}/3)$, where Greek indices denote Cartesian coordinates. Here, $\mathbf{n}$ is the director field representing the average molecular orientation, while the scalar $q$ quantifies the degree of local order. The magnitude $q$ is proportional to the largest eigenvalue of $\mathbf{Q}$ and is bounded by $0\leq q \leq 3/2$. 

The physical properties of the liquid crystal are captured by a free energy $\mathcal{F}$ that depends on the order parameter $\bQ$. The most common form for its density is the Landau de-Gennes free energy density
\begin{equation}
\begin{split}
    f_{LC} = &\frac{A_0}{2}\left(1-\frac{\gamma}{3}\right)Q_{\alpha\beta}^2 - \frac{A_0 \gamma}{3}Q_{\alpha\beta}Q_{\beta\gamma}Q_{\gamma\alpha} \\ &+ \frac{A_0\gamma}{4}(Q_{\alpha\beta}^2)^2
    + \frac{K}{2}(\partial_\alpha Q_{\beta\gamma})^2,
    \label{eq:f_LC}
\end{split}
\end{equation}
where Einstein summation convention is implied. The polynomial terms in Eq.~\eqref{eq:f_LC} account for the bulk free energy density, where $A_0$ is a positive constant that sets the energy scale. The parameter $\gamma$ acts as an effective inverse temperature as increasing $\gamma$ (cooling) drives the isotropic–nematic transition, with the nematic phase stable for $\gamma > \gamma_c = 2.7$. The final term accounts for the elastic energy associated with the director field. We adopt the one-constant approximation, setting the energy costs for splay, bend, and twist distortions as equivalent, and governed by the constant $K$.

The evolution of the $\bQ$ tensor is given by the Beris-Edwards equation~\cite{BerisEdwards1994}
\begin{equation}
        \left(\partial_t + \mathbf{v}\cdot\nabla \right) \bQ - \mathbf{S} = \Gamma\mathbf{H},
        \label{eq:evolution_Q}
\end{equation}
where $\Gamma$ is a collective rotational diffusion constant and $\bH$ is the molecular field, given by
\begin{equation}
       \mathbf{H}= -\frac{\delta F}{\delta \bQ}+\frac{\mathbf{I}}{3}\text{Tr}\frac{\delta F}{\delta \bQ},
       \label{eq:molecular_field}
\end{equation}
where $\mathbf{I}$ is the identity matrix and Tr denotes the tensorial trace. The first term in Eq.~\eqref{eq:evolution_Q} is the material derivative of $\bQ$ describing its time evolution under advection by a fluid with velocity $\bv$. The second term in Eq.~\eqref{eq:evolution_Q} describes how the molecules of the liquid fluid rotate and stretch in response to flow gradients~\cite{BerisEdwards1994}, defined explicitly as
\begin{equation}
\begin{split}
    \mathbf{S}=&(\xi\mathbf{D}+\boldsymbol{\omega})\cdot\left(\bQ + \frac{\mathbf{I}}{3}\right) + \left(\bQ + \frac{\mathbf{I}}{3}\right)\cdot \left(\xi\mathbf{D} -
    \boldsymbol{\omega}\right)\\
    &-2\xi\left(\bQ + \frac{\mathbf{I}}{3}\right)\text{Tr}(\bQ\cdot \nabla\mathbf{v)}.
    \label{eq:S_term}
\end{split}
\end{equation}
Here, $\mathbf{D}=(\nabla\mathbf{v}+\nabla\mathbf{v}^T)/2$ and $\boldsymbol{\omega}=(\nabla\mathbf{v}-\nabla\mathbf{v}^T)/2$ are the symmetric and antisymmetric parts of the velocity gradient tensor $\nabla \bv$, representing the rate of strain and vorticity, respectively. The tumbling parameter $\xi$ determines the molecular shape anisotropy, where positive values correspond to rod-like molecules and negative values to disk-like ones. The term on the right-hand side of Eq.~\eqref{eq:evolution_Q} describes the relaxation of the tensorial order parameter toward the free energy minimum.

Lastly, the fluid velocity evolves according to the Navier-Stokes equation
\begin{equation} \rho(\partial_t+u_\beta\partial_\beta)u_\alpha= - \partial_\alpha p_0 + \partial_\beta\Pi_{\alpha\beta}+\eta\partial^2_\beta u_\alpha,
    \label{eq:navier_stokes}
\end{equation}
under the incompressibility condition $\partial_\alpha u_\alpha=0$. In Eq.~\eqref{eq:navier_stokes} $\rho$ is the fluid density, $\eta$ is an isotropic viscosity and $p_0$ the hydrodynamic pressure. $\Pi_{\alpha\beta}$ is the stress that arises from the from liquid crystal deformations and is given by
\begin{equation}
\begin{split}
    \Pi_{\alpha\beta} = &2\xi\left(Q_{\alpha\beta}+\frac{1}{3}\delta_{\alpha\beta}\right)Q_{\gamma\epsilon}H_{\gamma\epsilon}-\xi\left(Q_{\gamma\beta}+\frac{1}{3}\delta_{\gamma\beta}\right)H_{\alpha\gamma} \\&- \xi\left(Q_{\alpha\gamma}+\frac{1}{3}\delta_{\alpha\gamma}\right)H_{\gamma\beta}
    -\partial_\alpha Q_{\gamma\nu}\frac{\partial f}{\partial (\partial_\beta Q_{\gamma\nu})} \\ &+ Q_{\alpha\gamma}H_{\gamma\beta} -H_{\alpha\gamma}Q_{\gamma\beta}.
\label{eq:stress_tensor}
\end{split}
\end{equation}

\subsection*{Binary Mixture}

The local composition of the binary mixture is given by the compositional order parameter $\phi$, such that $\phi=0$ corresponds to an isotropic fluid and $\phi=2$ to a pure nematic liquid crystal. The free energy density that describes a non-phase-separating mixture is given by 
\begin{equation}
    f_\phi = \frac{a}{2}\phi^2 + \frac{\kappa}{2}(\partial_\alpha\phi)^2 +WQ_{\alpha\beta}\partial_\alpha\phi\partial_\beta\phi,
    \label{eq:f_phi}
\end{equation}
where $a$ is a positive constant. The second term in Eq.~\eqref{eq:f_phi} penalizes spatial gradients of the order parameter, thereby defining the interfacial energy. The two terms $a$ and $\kappa$ scale the bulk and gradient contributions in the free energy, and  contribute to the surface tension and interface thickness between different emerging phases. %dictate the surface tension $\sigma\sim\sqrt{\kappa a}$ and the interface thickness $\xi_\phi\sim\sqrt{\kappa/a}$ between different phases. %Davide: slightly rewritten, as i agree this is only true for square well
The third term in Eq.~\eqref{eq:f_phi} couples the director field to compositional gradients, enforcing preferential anchoring at interfaces. The anchoring strength $W$ determines whether the anchoring is homeotropic ($W<0$) or planar ($W>0$).

Similarly to what was done in a previous work~\cite{AssanteEtAl2023} we define a linear dependence of $\gamma$ (from Eq.~\eqref{eq:f_LC}) on the local nematic composition
\begin{equation}
    \gamma=\gamma_0+\Delta \ \phi,
    \label{eq:gamma_def}
\end{equation}
where $\gamma_0$ is a bare coupling parameter and $\Delta$, the coupling parameter, a positive constant.

Finally, the evolution of the compositional order $\phi$ follows a Cahn-Hilliard equation
\begin{equation}
    \partial_t\phi+\mathbf{v}\cdot \nabla \phi = M\nabla^2\mu,
    \label{eq:cahn_hilliard}
\end{equation}
where $M$ is the mobility and $\mu=\delta F/\delta \phi$ is the chemical potential.

This work uses a 2D hybrid lattice Boltzmann (LB) scheme~\cite{Carenza2019_2}, solving Eq.~\eqref{eq:evolution_Q} and Eq.~\eqref{eq:cahn_hilliard} with finite differences, while Eq.~\eqref{eq:navier_stokes} was integrated using the LB approach. The fixed parameters used to generate the results of Fig.~\ref{fig:phase_diagrams} and the snapshots of Figs.~\ref{fig:snapshots},~\ref{fig:domain_growth} and ~\ref{fig:super_smectics} are, unless previously stated otherwise, $A_0=0.25$, $K=0.01$, $\Gamma=1$, $\xi=0.7$, $\eta=1.7$, $a=0.01$, $\kappa=0.01$ and $M=1$.

\subsection*{Topological charge density}
For our system, we compute the topological charge density $q_t$ as~\cite{blow2014}
\begin{equation}\label{qt}
    q_t(x,y) = \frac{1}{2\pi}\left(\partial_xQ_{xx}\partial_yQ_{xy}-\partial_xQ_{xy}\partial_yQ_{xx}\right).
\end{equation}
It is important to note that Eq.~(\ref{qt}) is well defined for a lyotropic mixture as well as a single-fluid nematic, because the ${\mathbf Q}$ tensor is defined at each point, and both in nematic and isotropic regions. As a result, because in our mixture the system can create droplets and voids, the topological charge of a region can delocalise or spread over interfaces rather than being localised at a singularities as normally occurs in a single-phase nematic. 

We note for completeness that $q_t$ is also the $zz$ component of the recently introduced  disclination tensor \cite{head2024,schimming2022,fnx3htyx}, 
\begin{align}\label{defDij}
    D_{ij}=\epsilon_{i\mu\nu}\epsilon_{jlk}\partial_l Q_{\mu\alpha}\partial_k Q_{\nu\alpha},
\end{align}
where $i,j,k,\alpha,\mu,\nu$ are tensor indices and where the Einstein summation convention of repeated indices has been used.

\subsection*{Preparation of Sunset Yellow Lyotropic Phase}
Sunset Yellow FCF (SSY) purchased from Sigma Aldrich, with purity $>$90\,\%, was further purified following the method of Horowitz et al.~\cite{janowitz2005}.
This involved dissolving SSY in Milli-Q water and adding ethanol, causing it to precipitate. 
The precipitate was then isolated with filter paper, and dried in a vacuum oven. 
This procedure was carried out twice before being used in experiments. 
Samples of lyotropic SSY were created by adding dried, purified SSY to Milli-Q water at 29\,wt\%, and heated at around 50\,$^\circ$C until the SSY was dissolved completely.

\subsection*{Temperature Ramp Experiments with Sunset Yellow}

\textcolor{black}{Lyotropic SSY was loaded into a 20 $\mu$m-thick cell, which was then sealed prior to temperature ramp experiments. Temperature ramps were performed using an Instec mK1000, with the temperature increased at rates of 0.1, 0.2, and 0.5 $^\circ$C min$^{-1}$.
For black-and-white time-lapse imaging, samples were observed using a Nikon ECLIPSE TE300 microscope, with movies recorded using a Mikrotron EoSens MC1362 high-speed camera. Colour still images were acquired using an Olympus BH2-UMA microscope, with images captured using a QImaging MicroPublisher 3.3 RTV camera.
All samples were viewed between crossed polarisers.}

%\showmatmethods{} % Display the Materials and Methods section
\section{Acknowledgments}
This research has received funding from the European Research Council (ERC) under the European Union’s Horizon 2020 research and innovation program (Grant Agreement No. 851196). This work used the Cirrus UK National Supercomputing Service through UKRI project ec302 (ENACT (Engineered Nematic Active Composites and Topology)).
For the purpose of open access, the author has applied a Creative Commons Attribution (CC BY) licence to any Author Accepted Manuscript version arising from this submission.

%\showacknow{} % Display the acknowledgments section

%\bibsplit[2]
%Use \bibsplit to split the references from the body of the text. Value "[2]" represents the number of reference in the left column (Note: Please avoid single column figures & tables on this page.)

% Bibliography
\bibliography{bib}

%\clearpage

\newpage
\onecolumngrid
\clearpage
\section{Supplementary Information}
\subsection{Identification of a phase separated regime}

To build the phase diagram from the simulations and identify the phase separated regimes, the average nematic order and the
Binder cumulants were computed. 

The local nematic order corresponds to the highest eigenvalue of the $\textbf{Q}$-tensor in each point in space. The average nematic order $\langle q \rangle$, is the average of the local one over all the points. In Fig.2B of the main text, $\langle q \rangle$ is plotted for different values of $\gamma_0$ and $\phi_0$. There is a clear transition from the isotropic phase ($\langle q \rangle \approx 0$) to the coexistence region ($0.2<\langle q\rangle <0.5$) and nematic ($\langle q\rangle \approx 2/3$).

The Binder cumulant is computed from
\begin{equation}
    U_\phi = 1-\frac{\langle \phi\rangle^4}{3\langle \phi^2\rangle^2},
    \label{eq:binder}
\end{equation}
where the averages are taken over all the points in space for a given configuration. According to~\eqref{eq:binder}, a Binder cumulant under 2/3 is indicative of a coexistence of different phases. In Fig.2C of the main text values of the Binder cumulant of $0.2<U_\phi<0.5$ confirm the coexistence of nematic and isotropic phases.

\subsection{Calculations for the common tangent construction}

Let us consider the free energy of the binary mixture given by Eq.(7) and Eq.(13) of the main text. The full free energy density is thus
\begin{equation}
    \begin{split}
      f = \frac{A_0}{2}\left(1-\frac{\gamma(\phi)}{3}\right)Q_{\alpha\beta}^2 &- \frac{A_0 \gamma(\phi)}{3}Q_{\alpha\beta}Q_{\beta\gamma}Q_{\gamma\alpha} \frac{A_0\gamma(\phi)}{4}(Q_{\alpha\beta}^2)^2
    + \frac{K}{2}(\partial_\alpha Q_{\beta\gamma})^2 \\ 
    &+ \frac{a}{2}\phi^2 + \frac{\kappa}{2}(\partial_\alpha\phi)^2 + WQ_{\alpha\beta}\partial_\alpha\phi\partial_\beta\phi.
    \end{split}
    \label{free-energy}
\end{equation}
To determine the phase boundaries of the system, we consider a homogeneous configuration of unvarying composition and
uniformly aligned nematic phase. In this state, the third (elastic energy), fifth (interfacial energy), and seventh (anchoring) terms vanish, and the homogeneous free energy density depends only on the nematic order $q$ and the composition $\phi$.

Given that the nematic order parameter is given by $Q_{\alpha\beta}=q(n_\alpha n_\beta-\delta_{\alpha\beta}/3)$, the homogeneous free energy simplifies to:
\begin{equation}
    f_\text{hom}(q,\phi) = \frac{A_0}{3}\left(1-\frac{\gamma(\phi)}{3}\right)q^2 - \frac{2 A_0\gamma(\phi)}{27}q^3+\frac{A_0\gamma(\phi)}{9}q^4+\frac{a}{2}\phi^2,
\end{equation}
where the total free energy is minimised for the nematic order parameter
\begin{equation} 
q_\text{min}= 
\begin{cases}
0 & \text{if } \gamma\leq \gamma_c, \\ 
\frac{1}{4}\left(1+\sqrt{9-24/\gamma(\phi)}\right) & \text{if } \gamma\geq\gamma_c,
\end{cases} 
\end{equation}
where $\gamma(\phi) = \gamma_0 + \Delta \phi$ and $\gamma_c=2.7$. This free energy has two concavities (inset of Fig. 2D of the main text), thus defining a region where the second derivative of the free energy is negative, and the mixture is unstable. 

The first derivative of the free energy is given by $\partial f_\text{hom}/\partial \phi = a\phi$ below $\gamma_c$ and above $\gamma_c$ by
\begin{gather*}
    \frac{\partial f_\text{hom}}{\partial \phi} =
-\frac{A_0\Delta}{9}q^2 - \frac{2 A_0\Delta}{27}q^3+\frac{A_0\Delta}{9}q^4+a\phi + \left[ \frac{2A_0}{3}\left(1-\frac{\gamma(\phi)}{3}\right)q - \frac{2 A_0\gamma(\phi)}{9}q^2+\frac{4A_0\gamma(\phi)}{9}q^3\right]q'.
\end{gather*}
The second derivative is $\partial^2 f_\text{hom}/\partial \phi^2=a$ below $\gamma_c$ and above $\gamma_c$ is 
\begin{gather*}
    \frac{\partial ^2 f_\text{hom}}{\partial \phi^2} = a + 2\left[-\frac{2A_0\Delta}{9}q - \frac{2 A_0\Delta}{9}q^2+\frac{4A_0\Delta}{9}q^3\right]    q' 
    + \left[ \frac{2A_0}{3}\left(1-\frac{\gamma(\phi)}{3}\right) - \frac{4 A_0\gamma(\phi)}{9}q+\frac{12A_0\gamma(\phi)}{9}q^2\right](q')^2
    \\
    + \left[ \frac{2A_0}{3}\left(1-\frac{\gamma(\phi)}{3}\right)q - \frac{2 A_0\gamma(\phi)}{9}q^2+\frac{4A_0\gamma(\phi)}{9}q^3\right]q'',
\end{gather*}
where 
\begin{align}
    \frac{\partial q}{\partial \phi} =\frac{3\Delta}{\gamma(\phi)^ 2}\left(9-\frac{24}{\gamma(\phi)}\right)^{-1/2}, && \frac{\partial ^2q}{\partial \phi^2} = \frac{-54 \Delta^2 (\gamma(\phi) - 2)}{(\gamma(\phi))^4 \left( 9 - \frac{24}{\gamma(\phi)} \right)^{3/2}}.
\end{align}

The phase diagram is built by varying the total composition of the mixture $\phi$, and the bare coupling parameter $\gamma_c$ and computing the respective spinodal and binodal regions numerically. 

The spinodal region is defined by $\partial^2 f_\text{hom}/ \partial\phi^2<0$ and is delimited by the points where the $\partial^2 f_\text{hom}/ \partial\phi^2$ is discontinuous, $\phi_c = (2.7-\gamma_0)/\Delta$, and where $\partial^2 f_\text{hom}/ \partial\phi^2=0$. The later point is estimated greedily by scanning through all the values of $\partial^2 f_\text{hom}/ \partial\phi^2$ above the discontinuity and the selecting the value of $\phi$ that yields the $\partial^2 f_\text{hom}/ \partial\phi^2$ closer to zero. 

The binodal region is found by a greedy iterative method. Firstly a fine grid of $\phi$ is defined, and those grid points are split into two intervals for $\gamma<\gamma_c$ and $\gamma>\gamma_c$. Then, all pairs $(\phi_1,\phi_2)$ formed by points of both intervals are sampled and the pair that satisfies the following condition 
\begin{equation}
 \min_{\{\phi_1,\phi_2\}}  \left\{\bigg\rVert \frac{\partial f_\text{hom}}{\partial \phi}\bigg\rvert_{\phi_1} - \frac{\partial f_\text{hom}}{\partial \phi}\bigg\rvert_{\phi_2} \bigg\rVert + \bigg\Vert g(\phi_1)-g(\phi_2)\bigg\Vert\right\},
 \label{condition}
\end{equation}
is found. In Eq.~\eqref{condition}, $g(\phi) = f_\text{hom}(\phi)-f_\text{hom}'(\phi)\phi$ is the $y$-intercept of the line tangent to the point $(\phi,f_\text{hom}(\phi))$. The pair that satisfies Eq.~\eqref{condition} defines the limits of the binodal region. The tangent line that connects the points $(\phi_1,f_\text{hom}(\phi_1))$ $(\phi_2,f_\text{hom}(\phi_2))$ represents the average free energy density of the phase-separated mixture, which is lower than that of the homogeneous state for any given average composition $\langle \phi\rangle$ lying between the binodals. 

\subsection{Critical anchoring}

Let us again consider the free energy density of Eq.~\eqref{free-energy}. To determine the anchoring strength for which phase separation is arrested, we examine a single lamellar nematic domain with a well-defined interface. We consider a region where the director field is spatially uniform ($\mathbf{n}(\mathbf{r})=\mathbf{n}$) and coupled to the
domain interface, such that $\mathbf{n}\cdot \nabla \phi /\Vert \nabla\phi\Vert=0$ for planar anchoring and $\mathbf{n}\cdot \nabla\phi /\Vert \nabla \phi\Vert=1$ for homeotropic anchoring. Under these conditions, the free energy simplifies to
\begin{equation}
    \begin{split}
      f_\text{lam} = \frac{A_0}{3}\left(1-\frac{\gamma(\phi)}{3}\right)q^2& - \frac{2 A_0\gamma(\phi)}{27}q^3+\frac{A_0\gamma(\phi)}{9}q^4
    + \frac{K}{3}\vert \nabla q\vert^2 + \frac{a}{2}\phi^2 + \frac{\kappa}{2}\vert\nabla\phi\vert^2 + Wq \left[|\mathbf{n}\cdot\nabla\phi|^2 - \frac{1}{3}\vert \nabla \phi\vert^2\right]
    \end{split}
    \label{free-energy}
\end{equation}
Assuming that the local nematic order is linearly coupled to the local concentration via $q(\mathbf{r}) \approx \alpha\phi(\mathbf{r})$, the free energy density can be recast to isolate the gradient contributions
\begin{equation}
\begin{split}
f_\text{lam} \approx f_\text{bulk}(\phi, q) + \kappa_\text{eff}(\phi, q) (\nabla \phi)^2,
\end{split}
\label{eq:f_lam}
\end{equation}
where an effective surface tension coefficient
\begin{equation}
    \kappa_\text{eff} = \frac{\kappa}{2} - \frac{W q}{3} + \frac{K \alpha^2}{3} + Wq \frac{\mathbf{n} \cdot \nabla \phi}{|\nabla \phi|}
    \label{eq:sigma_eff}
\end{equation}
appears. As described in the main text, macrophase separation is driven by a positive surface tension ($\kappa_{\text{eff}} > 0$), which drives the system to minimise interfacial area. However, the second term in $\kappa_{\text{eff}}$ (Eq.\eqref{eq:sigma_eff}) is negative and so a transition occurs when this anchoring contribution renders $\kappa_{\text{eff}}$ negative. The critical threshold for this transition ($\kappa_{\text{eff}} = 0$) is defined by
\begin{equation}
    \left|W_c^\perp\right| = \frac{3}{2q}\left(\frac{\kappa}{2}+\frac{K\alpha^2}{3}\right), \quad
    % W_c^\parallel = \frac{3}{q}\left(\frac{\kappa}{2}+\frac{K\alpha^2}{3}\right)
    W_c^\parallel = 2 \left|W_c^\perp\right|,
    \label{eq:Wc_analytical}
\end{equation}
for homeotropic and planar anchoring, respectively.

\subsection{Super-smectics length scales characterisation}

The lamellar patterns that emerge for high values of anchoring strength were characterised by measuring the oriented correlation function of the composition $\phi$, and the structure factor. 

The oriented correlation function is given by
\begin{equation}
    G(\Delta r ) = \frac{\langle\phi(\mathbf{r}+\mathbf{\hat{e}}\Delta r)\phi(\mathbf{r})\rangle}{\langle \phi\rangle},
\end{equation}
where the averages are spatial averages, taken over all the points in a given configuration, and $\mathbf{\hat{e}}$ is a unit versor. For the characterisation of the lammellar patterns shown in Fig.~\ref{fig:correlation_structure_factor}, the correlation function was sampled along the $y$ direction (denoted as $G(y)$) and along the direction perpendicular to the \textit{lamellae} (denoted as $G(r_\perp)$). 

In panels A and B of Fig.\ref{fig:correlation_structure_factor}, $G(y)$ is shown for three specific configurations with homeotropic (pink lines), planar (blue lines) and no anchoring (bronze lines). Given that for these configurations the emergent lamellar patterns tendentially align themselves perpendicular to the $y$ axis, the correlation function was sampled along $y$. In the absence of anchoring, $G(y)$ decays rapidly. In contrast, the oriented correlation functions for strong homeotropic and planar anchoring show a distinct decaying oscillatory trend. The high frequency oscillations are marks of the lamellar structure, with a wavelength of approximately $5<\lambda^*<8$ lattice units. In the case of homeotropic anchoring, there is not only a single decay in the correlation, but also an envelope whose wavelength ($\lambda \approx 50$) is related to the characteristic size of the domains of \textit{lamella} that align in a single direction. The detection of an enveloping frequency for homeotropic anchoring is a mark of the more well defined structures that are found in this case in comparison to the planar anchroing case. A further attempt at quantifying the length scales of the system by sampling the correlation function along the direction perpendicular to the \textit{lamellae}. These values are plotted for the systems present in the inset of Fig.~\ref{fig:correlation_structure_factor}B in the presence of homeotropic (pink) and planar (blue) anchoring in Fig.~\ref{fig:correlation_structure_factor}F-H. The results are similar to those of Fig.~\ref{fig:correlation_structure_factor}B, with rapidly decaying oscillations for planar anchoring and enveloped ones for homeotropic anchoring. Panels G and H show semi-log plots of $|1-G(r_\perp)|$ for homeotropic and planar anchoring respectively. From the distance between the peaks in Fig.~\ref{fig:correlation_structure_factor}G one can take $\lambda^*\approx 6$. However, as seen in the insets of panels A and B, though the \textit{lamella} thickness is well defined, the spacing between \textit{lamellae} is not. This coupled with the fact that the \textit{lamella} orientation is not spatially uniform makes it difficult to quantify precisely the length scales that are present in the system. 

These length scales were further quantified via the structure factor (see Fig.~\ref{fig:correlation_structure_factor}C-E), computed as
\begin{equation}
    S(\mathbf{k})= \bigg\rvert \int \text{d}^2\mathbf{r} e^{-i\mathbf{k}\cdot \mathbf{r}}(\phi(\mathbf{r})-\langle \phi(\mathbf{r})\rangle ) \bigg\rvert ^2.
\end{equation}
Peaks emerge for frequencies of approximately $0.5 < k < 1.2$, which correspond to wavelengths of $5.2 < \lambda < 13$. These peaks are clearer for homeotropic anchoring (Fig.~\ref{fig:correlation_structure_factor}C) in which the lamellar structure is better defined. In contrast, for planar anchoring the peaks are more disperse and there is a bounding frequency for $k\approx 1.2$ that relates to the finer lengths scales in the system. Since the \textit{lamella} are less pronounced in this case, we can also identify peaks at lower frequencies $k\approx 0.15 \implies \lambda\approx 42$ that relate to the domain sizes in which the \textit{lamella} are pointing in the same direction. In the absence of anchoring, the structure factor peaks relate to the droplet size.

%\textcolor{red}{ANA: Even tough agreed to remove Fig 3. I wrote the section on the characterisation in case we need to show it.}

\subsubsection{Detecting critical anchoring}

As explained in the main text, the critical anchoring, that is, the anchoring strength from which lamelar pattern formation occurs and phase separation is arrested, the phase space of the surface tension coefficient $\kappa$ and the nematic elastic constant $K$ was explored (values below). For each pair $(\kappa,K)$, the anchoring was gradually increased until the patterns were detected. 

To detect the onset of pattern formation the structure factor was considered. As seen above (Fig.~\ref{fig:correlation_structure_factor}E), when planar anchoring is considered, the structure factor also exhibits pronounced peaks at lower frequencies, which are reminiscent of larger scale structures. Therefore, to detect the presence of the finer laminar structures, a mask was first applied to remove the lower frequency peaks (corresponding to $\lambda>8$). Then, the maximum peak of the structure factor was computed, and for $|S(\mathbf{{k}})|>10^4$ we considered patterns to have formed. The first value of $W$ satisfying this condition was called the critical anchoring strength $W_c$ and used for the linear fit.

Values considered:

\begin{itemize}
    \item $\kappa \in \{0.001, 0.002, \cdots, 0.014\}$
    \item $K \in \{0.005, 0.006, \cdots, 0.034\}$
    \item $\vert W^\perp\vert  \in \{0.001, 0.001, \cdots, 0.045\}$
    \item $\vert W^\parallel\vert  \in \{0.001, 0.001, \cdots, 0.085\}$
\end{itemize}

By performing a bilinear fit of the obtained values of $W_c^\perp$ and $W_c^\parallel$ as functions of $\kappa$ and $K$, the following fits are obtained:

\begin{align}
    \vert W_c^\perp| = 2.001\kappa + 0.138 K + 0.008 \quad (\pm 0.001)\\
    W_c^\parallel = 3.987\kappa + 0.253 K + 0.014 \quad (\pm 0.001)
    \label{eq:fit-anchoring}
\end{align}

and by forcing a null $y-$intercept:
\begin{align}
    \vert W_c^\perp| = 2.401\kappa + 0.364 K  \quad (\pm 0.003)\\
    W_c^\parallel = 4.681\kappa + 0.644 K \quad (\pm 0.005)
    \label{eq:fit-anchoring}
\end{align}

\subsection{Supplementary Figures}
\begin{figure}[H]
    \centering
    \includegraphics[width=0.75\linewidth]{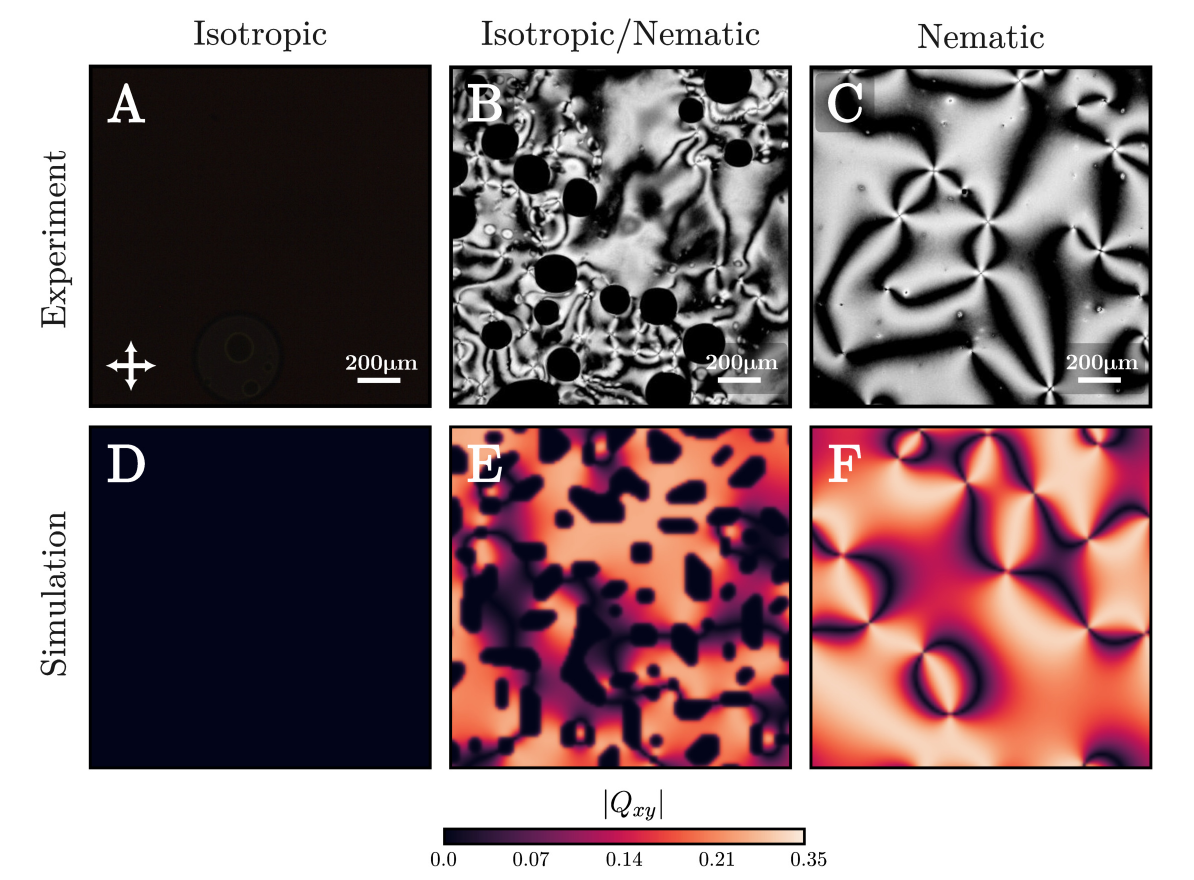}
    \caption{
(A–C) Experimental images showing the isotropic (A), coexistence (B), and nematic (C) phases as the temperature decreases ($\gamma_0$ increases) and the nematic fluid concentration $\phi_0$ increases. (D–F) Corresponding simulation snapshots illustrating the same sequence of phase behaviour and Schlieren patterns for a 128$\times$128 system. The colour map represents the modulus of the $\mathbf{Q}$-tensor component $|Q_{xy}|$. Scale bars: \SI{200}{\micro\meter}.}
    \label{fig:fig1_Qxy}
\end{figure}
\begin{figure}[H]
    \centering
    \includegraphics[width=0.75\linewidth]{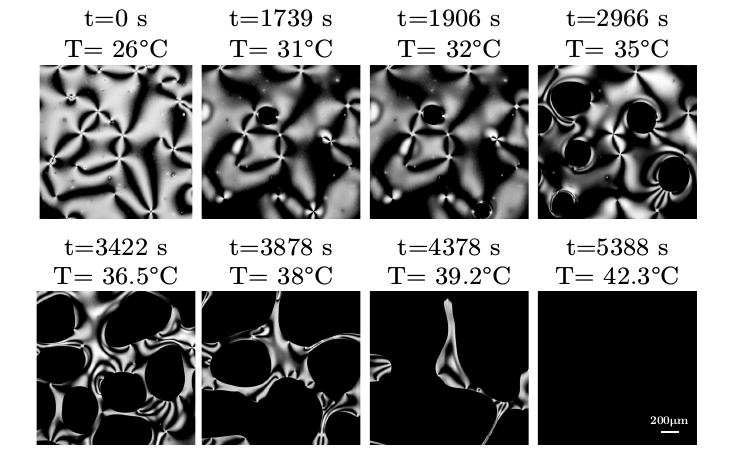}
    \caption{Transition from a fully nematic to fully isotropic phase of a mixture of SSY at 28 wt$\%$ undergoing a temperature ramp from 26.0 to 42.3 $^\circ$C. Snapshots taken from Movie 1.}
    \label{fig:temp_ramp_fig3}
\end{figure}

 %\begin{figure}[!h]
  %  \centering
   %  \includegraphics[width=0.7\textwidth]{figures_SI/exp_coarsening_temp_ramp.pdf}
   %  \caption{Transition from a fully nematic to fully isotropic phase of a mixture of SSY at 28 wt$\%$ undergoing a temperature ramp from 26 to 43 $^\circ$C. Scale bar \SI{200}{\micro\meter}. Snapshots taken from Movie 1.}
   %  \label{fig:temp_ramp}
 %\end{figure}  

\begin{figure}[!h]
    \centering
    \includegraphics[width=\linewidth]{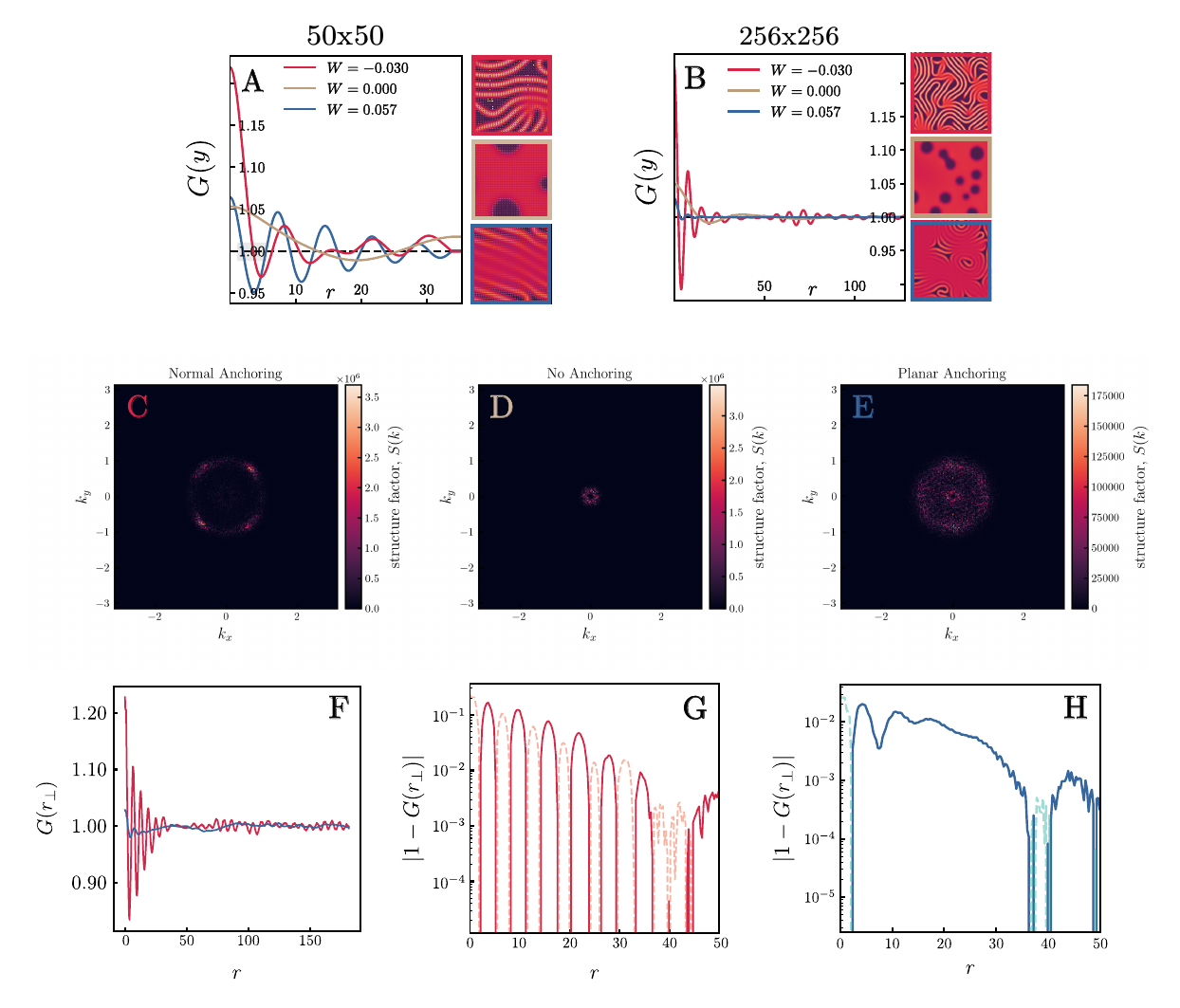}
    \caption{Structural analysis of the lamellar patterns with planar, homeotropic and zero anchoring. For all simulations $\phi_0=1.0$, $\gamma_0=2.0$ and $\kappa=0.01$. A and B show the spatial correlation function for the nematic composition $\phi$ sampled in the $y$ direction, $G(y)$, for homeotropic ($W=-0.030$), planar ($W=0.057$) and zero anchoring $W=0$, for two different system sizes. Side panels show the corresponding configurations. Panels C-E show the corresponding structure factor for the largest system size. Panel F shows the spatial correlation function sampled in direction parallel to the lamellar normals $G(r_\perp)$, for homeotropic (pink) and planar (blue) anchoring. G and H show the respective curves for $|1-G(r_\perp)|$ (semi-log).}
   \label{fig:correlation_structure_factor}
\end{figure}

\newpage
\subsection{Movies}

\begin{itemize}
    \item Movie 1 - Phase separation and coarsening starting from a random, noisy initial configuration with no surface tension ($\kappa=0$) or anchoring ($W=0$). Zoomed out animation of the full 128$\times$128 system presented in Fig.3A of the main text. The colour map represents the local compositional phase, $\phi$, while the overlaid lines indicate the nematic director field, with line length proportional to the local degree of orientational order. Global compositional phase is fixed at $\phi_0=1.0$ and the bare coupling parameter at $\gamma_0=2.0$. As time passes, defects annihilate, irregular droplets form and domains coarsen over time due to Ostwald ripening.
    \item Movie 2 - Phase separation and coarsening starting from a random, noisy initial configuration with surface tension ($\kappa=0.01$) and no anchoring ($W=0$). Zoomed out animation of the full 128$\times$128 system presented in Fig.3B of the main text. The colour map represents the local compositional phase, $\phi$, while the overlaid lines indicate the nematic director field, with line length proportional to the local degree of orientational order. Global compositional phase is fixed at $\phi_0=1.0$ and the bare coupling parameter at $\gamma_0=2.0$. As time passes isotropic droplets form from the annihilation of topological defects. Droplets are rounder in shape and coalesce, in resemblance to the experiments.
    \item Movie 3 - Mixture of SSY at 28 wt$\%$ undergoing a temperature ramp from 26 to 43 $^\circ$C. Snapshots from this movie are included in Fig.3 of the main text and Supp. Fig.~\ref{fig:temp_ramp_fig3}. As the temperature increases, defects annihilate and isotropic droplets emerge and coarsen over time until eventually all of the mixtures becomes isotropic. 
    \item Movie 4 - Mixture of SSY at 28 wt$\%$ undergoing a temperature ramp from 25 (nematic phase) to 45 $^\circ$C (isotropic phase). In the coexistence regime the nematic phase forms \textit{lamella} structures of well defined thickness. A snapshots from this movie is included in Fig.4A of the main text.
    \item Movie 5 - Time evolution of a simulation with $\phi_0=1.0$, $\gamma_0=2.0$, $\kappa=0.01$ and $W=-0.03$ (homeotropic anchoring) starting from a random and noisy configuration. Defect dynamics are shown: $+1/2$ (yellow) and $-1/2$ (blue) defects appear, move, and annihilate to relax the lamellar structure. However, full annihilation does not occur; anchoring stabilises microphase separation. This movie relates to panel D from Fig.4 of the main text, in this case, defects are shown reappearing transiently. System size: 50 $\times$ 50.
    \item Movie 6 - Time evolution of a simulation with $\phi_0=1.0$, $\gamma_0=2.0$, $\kappa=0.01$ and $W=-0.03$ (homeotropic anchoring) starting from a random and noisy configuration. Defect dynamics are shown: $+1/2$ (yellow) and $-1/2$ (blue) defects appear, move, and annihilate to relax the lamellar structure. However, full annihilation does not occur; anchoring stabilises microphase separation. This movie relates to panel E from Fig.4 of the main text, in this case, defects are shown to act as a source of the nematic layers. System size: 50 $\times$ 50.
    \item Movie 7 - Simulation initialised with a horizontally aligned lamellar pattern subject to additive noise. The initial bands are uniformly distributed throughout the domain, with widths comparable with those observed in Fig. 4B and Supplementary Movies 5 and 6 ($\lambda^* \approx 6$). The stability of the \textit{lamella} width indicates that the system has reached an equilibrium state. Simulation parameters correspond to those of Fig. 4B: surface tension $\kappa=0.01$, homeotropic anchoring $W=-0.03$, global composition $\phi_0=1.0$, and bare coupling constant $\gamma_0=1.0$.
    \item Movie 8 - Horizontal lamellar pattern with heterogeneous spacing and additive noise. Band widths correspond to those of Fig. 4B ($\lambda^* \approx 6$). The persistence of unoccupied regions highlights the unusually low layer-compression modulus of the self-assembled smectics. Simulation parameters correspond to those of Fig. 4B: surface tension $\kappa=0.01$, homeotropic anchoring $W=-0.03$, global composition $\phi_0=1.0$, and bare coupling constant $\gamma_0=1.0$.
    \item Movie 9 - Simulation initialised with a diagonal lamellar pattern subject to additive noise. The initial structures evolve into thinner lamellae with widths comparable to those observed in Fig. 4B, D and E, and Supplementary Movies 5 and 6. Simulation parameters correspond to those of Fig. 4B: surface tension $\kappa=0.01$, homeotropic anchoring $W=-0.03$, global composition $\phi_0=1.0$, and bare coupling constant $\gamma_0=1.0$.
\end{itemize}

All the movies are available \href{https://www.dropbox.com/scl/fo/kgbkz7m77aitec8v6o1w3/AEYdAjn9AGDvLvd-twxK7_E?rlkey=n8b65b2mr7a6i9wi8dss4xs13&e=1&st=anvsc8rk&dl=0}{here}.

\end{document}